\input{aipcheck}

\documentclass[
    ,final            
  ]
  {aipproc}

\layoutstyle{8x11double}

\begin{document}

\title{Dynamics of rotating spirals in agitated wet granular matter}


\author{Kai Huang}{
address={Experimentalphysik V, Universit\"at Bayreuth, 95440 Bayreuth, Germany}
}

\author{Lorenz Butzhammer}{
}

\author{Ingo Rehberg}{
}

\begin{abstract}
Pattern formation of a thin layer of vertically agitated wet granular matter is investigated experimentally. Due to the strong cohesion arising from the capillary bridges formed between adjacent particles, agitated wet granular matter exhibits a different scenario compared with cohesionless dry particles. Rotating spirals with three arms, which correspond to the kinks between regions with different colliding phases with the vibrating plate, have been found to be the dominating pattern 
\cite{Huang11}. From both top view snapshots and laser profilometry methods, the rotation frequency of the spiral arms is characterized with image processing procedures. Both methods reveal that there exists a finite rotation frequency $\nu_{\rm r}$ at a threshold vibration acceleration, above which $\nu_{\rm r}$ increases linearly with the peak vibration acceleration with a slope strongly dependent on the vibration frequency. 
\end{abstract}

\maketitle


From natural phenomena such as the formation of our galaxy, to the industrial processes such as coating, pattern formation is ubiquitous in the universe \cite{Cross09}. Explaining pattern formations, especially the accompanying length and time scales, helps us gain insight into the pattern forming systems, which are typically far from thermal equilibrium. Driven granular matter, as large agglomerations of macroscopic particles interacting with each other dissipatively, is such a nonequilibrium system \cite{Nagel96}. Continuous energy injection is necessary in order to drive granular particles into a fluidlike state in order to process or transport them in various industries \cite{Duran00}. Even though a all-encompassing theory describing the dynamical behavior of granular matter, especially toward the limit of dense granular flow, is still far from established, the rich pattern forming nature of granular matter provides us the opportunity to gain insights into such a nonequilibrium system \cite{Aranson06}.

Besides the cohesionless dry granular matter, more and more attentions have been paid to the dynamics of wet granular matter recently \cite{Herminghaus05}. Due to the cohesion arising from the formation of liquid bridges between adjacent particles, the static as well as dynamical behavior of wet granular matter, such as the wet sand used for sculptures, differs from its dry counterpart dramatically \cite{Scheel08, Huang09l}. A former investigation on the pattern formation of vertically agitated wet granular layer reveals a peculiar period tripling spiral pattern, typically composed of 3 arms that rotate continuously \cite{Huang11}. The preferred number of arms arise from period tripling of the agitated granular layer, which breaks the symmetry and drives the rotation of spiral arms. In the current work, the rotation frequency of the spiral arms is characterized quantitatively with two methods: One based on the horizontal motion of the arms from the top view snapshots, and the other one based on the fluctuations of the surface profile from the laser profilometry method.

\begin{figure}
  \includegraphics[width=.35\textwidth]{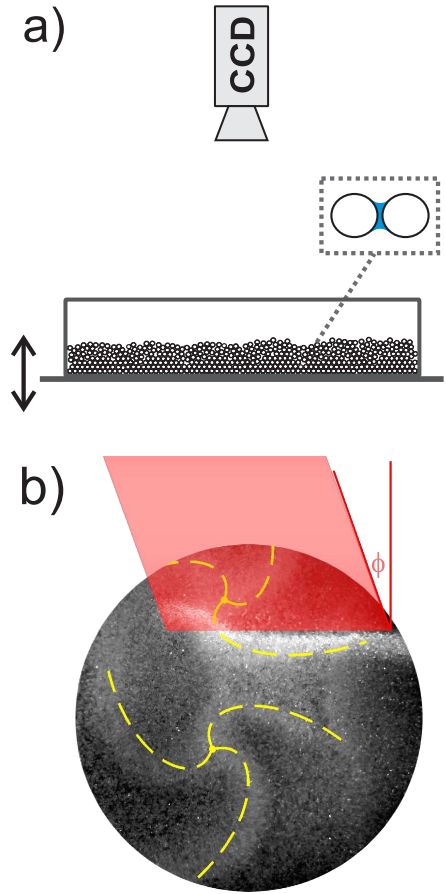}
  \caption{\label{setup} (a) A sketch of the experimental setup to capture the spiral pattern in a thin layer of wet granular matter under vertically sinusoidal agitations. The zoom-in view shown in the dotted box illustrates a capillary bridge formed between two adjacent particles. (b) A top view image captured with a combined illumination of a laser sheet and a LED ring. $\phi$ is the angle between the laser sheet (illustrated as half transparent red) and the horizontal plane. Yellow dash curves highlight the spiral arms.}
\end{figure}

Figure~\ref{setup}(a) is a sketch of the main parts of the experimental setup. Cleaned glass beads (SiLiBeads S) with a diameter of $d=0.78$\,mm and 10\% polydispersity, after proper mixing with purified water (Laborstar TWF-DI), are used as the granular sample. The amount of wetting is characterized by the liquid content $W=V_{\rm w}/V_{\rm g}$, where $V_{\rm w}$ is the volume of the water and $V_{\rm g}$ is that of the glass beads. It is normally kept within a few percent so that the cohesion arises mainly from the formation of capillary bridges between adjacent particles (as illustrated in the dash box). A certain amount of the sample (mass $m=113$\,g, corresponding roughly to 4--5 layers) is filled into a cylindrical polycarbonate container with an inner radius $R=8$\,cm and a height of $H=1.06$\,cm. Sinusoidal vibrations generated by an electromagnetic shaker (Tira TV50350) are used to agitate the sample against gravity. The vibration frequency $f$ and the nondimensional acceleration $\Gamma=4\pi^2 f^2 A/g$, with vibration amplitude $A$ and gravitational acceleration $g$, are the two control parameters. The mobility of the patterns are captured by a high speed camera (IDT MotionScope M3) mounted on top of the container. The camera is synchronized with the shaker by a multi-pulse generator, in order to capture images at fixed phases of each vibration cycle. A more detailed description of the experimental setup could be found elsewhere \cite{Huang11}.

\begin{figure}
  \includegraphics[width=.33\textwidth]{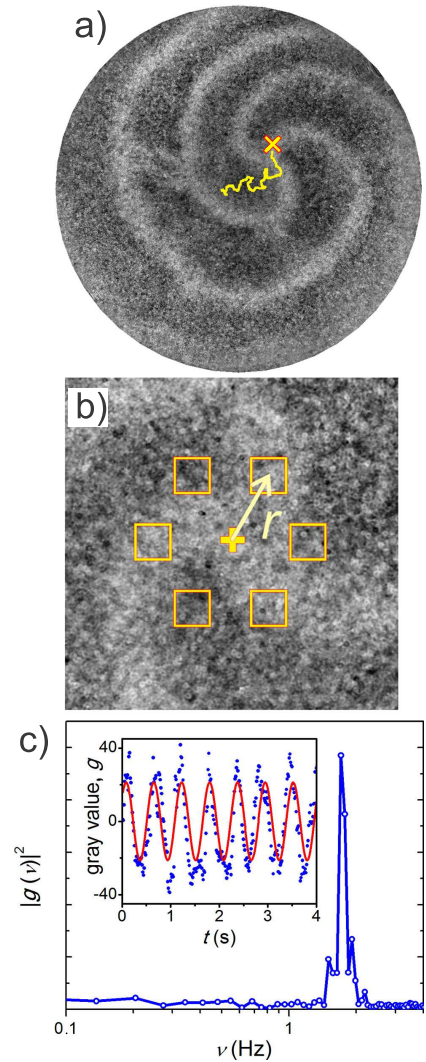}
  \caption{\label{top} Procedure to obtain rotation frequency from top view snapshots. (a) A typical image of a three armed rotating spiral pattern (averaged over 3 consequent snapshots to enhance the contrast). The yellow cross is the current position of the spiral core determined by the image processing procedure, and the yellow curve corresponds to the trajectory of the spiral core in the past 12.5 seconds (1000 vibration cycles). (b) A close view of the spiral core region. The regions used to obtain gray value fluctuations marked with squares have equal distance $r$ to the spiral core. (c) Averaged power spectrum $|g(\nu)|^2$ of the gray value fluctuations $g(t)$ obtained within 6 regions. Inset is the fluctuation with time (blue dots) and a fitting with a harmonic function (red curve).}
\end{figure}  

Two methods are used to capture the mobility of the spiral arms. The first one uses the snapshots of the pattern obtained with a low angle illumination, which provides the information on the horizontal motion. The second one uses the the laser profilometry method \cite{Raton95s}, which provides information on the surface height fluctuations. As sketched in Fig.~\ref{setup}(b), the laser sheet illuminates a line of the sample surface with a fixed angle $\phi\approx37^{\circ}$, which transfers the height profile of the illuminated line into the images plane.

From the top view images, the rotation frequency of the spiral arms can be obtained from the gray value fluctuations $g(t)$ of regions with fixed distance to the core of the spirals, based on the fact that the spiral arms in the snapshots are brighter than the other regions, as Fig.~\ref{top}(a) shows. The gray value difference is due to the fact that spiral arms correspond to kinks separating regions with different heights \cite{Huang11}, and more light from the low angle illumination will be reflected to the camera in the kink (spiral arm) regions. 


\begin{figure}
  \includegraphics[width=.38\textwidth]{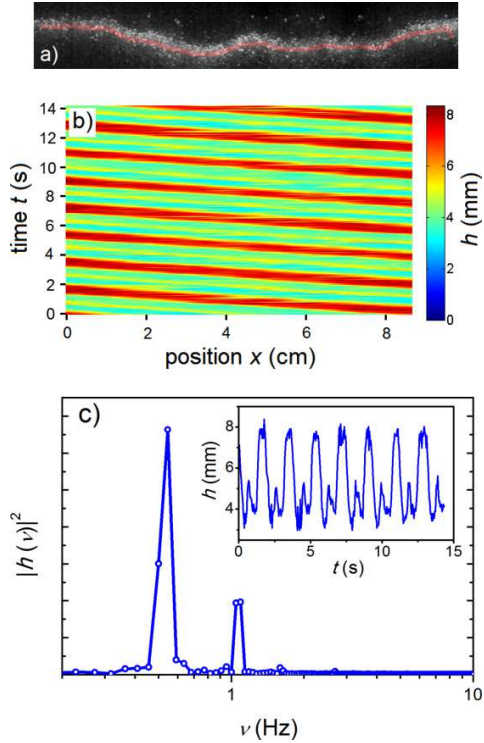}
  \caption{\label{side} Procedure to obtain rotation frequency from laser profilometry. (a) A snapshot captured with laser sheet illumination, from which the surface profile (curve with half transparent red) is obtained by image processing. (b) Time-space plot of the surface profile captured with a frame rate $f/3$. $h$ denotes the height of the profile. (c) Power spectrum of the height fluctuations $|h(\nu)|^2$, averaged over 6 positions with equal distances. The inset shows the height fluctuation with time at one position.}
\end{figure}

The essential part of the algorithm is to choose the region of interest for the analysis of the brightness fluctuations properly. Especially the regions close to the spiral cores should be avoided, since those regions correspond to the common part of spiral arms, where the brightness fluctuations are much smaller than those further away from the core. Therefore, our first step is to locate the spiral cores. Based on the fact that the spiral core region keeps its brightness during the rotation of spiral arms, a minimization of the gray value over continuous snapshots effectively enhances the contrast between the spiral core and the surrounding regions. The number of snapshots is chosen to cover $1/3$ of the rotation period. After proper thresholding, we are able to estimate the location of the spiral core from the weighted average of the core region. 

From the trajectory of the spiral cores determined (see the superimposed curve in Fig.~\ref{top}(a)), no clear relation to the rotation direction of the spiral arms could be found and the core wander around randomly. As shown in Fig.~\ref{top}(b), 6 square boxes with a certain radial distance $r$ to the spiral core, and equally distributed in the angular direction, are chosen as the regions of interest. The spatially averaged brightness within each box as a function of time is recorded individually (see the inset of Fig.~\ref{top}(c) for an example). As shown in Fig.~\ref{top}(c), the averaged power spectrum of the brightness over all the 6 boxes indicates clearly a periodicity of the gray value fluctuations. From the peak value ($|g(\nu_{\rm p0})|^2$ at frequency $\nu_{\rm p0}$) and the largest nearest neighbor value ($|g(\nu_{\rm p1})|^2$ at frequency $\nu_{\rm p1}$), we interpolate the rotation frequency with a weighted average $\nu_{\rm r}=(|g(\nu_{\rm p0})|\cdot\nu_{\rm p0}+|g(\nu_{\rm p1})|\cdot\nu_{\rm p1})/3(|g(\nu_{\rm p0})|+|g(\nu_{\rm p1})|)$, where the factor 3 arises from the fact that all the three spiral arms will generate brightness fluctuations. Alternatively, the frequency of $g(t)$ is also determined directly by a sinusoidal fit of the data in the time domain (red curve in the inset of Fig.~\ref{top}(c)), which agrees with the $\nu_{\rm r}$ from frequency domain within an error of $1$\%.

With the laser sheet illumination, the rotation of spiral arm is represented by the kink propagation along the illuminated line. Therefore the rotation frequency of the spiral arms corresponds to the repetition rate of the kink propagation. As shown in Fig.~\ref{side}(b), the time-space plot of the height profile clearly demonstrates this periodicity. The height profile of every third vibration cycle is recorded, in order to avoid the influence from the frequency of the surface height fluctuations ($f/3$) owing to period tripling. At a certain position, the plateau with maximal height (red color), which represents a region bounded by two spiral arms, repeatedly appears. The inset of Fig.~\ref{side}(c) shows one example of the height fluctuations with time $h(t)$. Following the above analysis on the $g(t)$, the rotation frequency is obtained from the power spectrum of the height fluctuations average over 6 equally distributed positions (see Fig.~\ref{side}(c)). Note that the height fluctuation is closer to a Heaviside step function than to a sinusoidal fluctuation, thus fitting directly on $h(t)$ is not appropriate. The smaller peak at about $2\nu_{\rm r}$ arises from the height fluctuation induced by the third spiral arm. 

\begin{figure}
  \includegraphics[width=.42\textwidth]{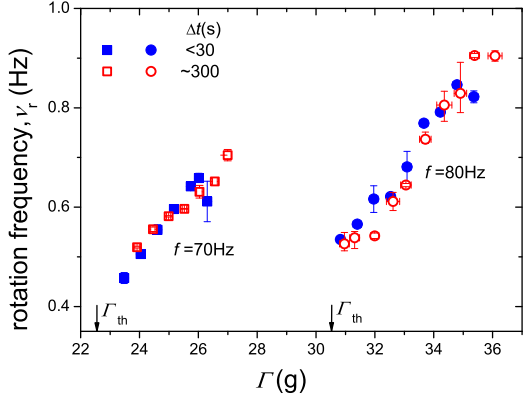}
   \caption{\label{time} Rotation frequency of the spiral arms $\nu_{\rm r}$ as a function of the peak vibration acceleration $\Gamma$ for two vibration frequencies ($f$) with liquid content $W=1.5\%$. The threshold for spiral pattern to emerge ($\Gamma_{\rm th}$) is $22.5\pm1.3$ for $f=70$\,Hz and $30.5\pm1.5$ for $f=80$\,Hz. The rotation frequencies are obtained from the top view snapshots of the pattern, which are captured either shortly after varying $\Gamma$ (with a delay time $\Delta t<30$\,s) or after a delay of $\Delta t\approx300$\,s.}
\end{figure}

Utilizing both methods, we are now ready to characterize the dynamics of rotating spirals. Qualitative observations reveal that the emerging spiral arms as the threshold acceleration is reached will have a finite rotation frequency. Our main task is to characterize the dependence of the rotation frequency on various control parameters. Here we focus only on the case of single spiral core and leave the case of multiple spirals for further investigations, because the interactions between spiral arms linked to different cores may lead to a variation of the rotation frequency.

Figure~\ref{time} shows the rotation frequency $\nu_{\rm r}$ as a function of $\Gamma$ for two vibration frequencies obtained from the top view snapshots taken at different time: shortly after changing control parameter $\Gamma$ or with a delay of $\Delta t \approx 300$\,s. For both vibration frequency $f$ and $\Gamma$, $\nu_{\rm r}$ captured at different time agree with each other within the errorbar, suggesting that the rotation speed of the arms is time independent for a constant driving condition. As $\Gamma$ is beyond the threshold value $\Gamma_{\rm th}$, the rotation frequency increases monotonically with $\Gamma$ within a certain range. Beyond the upper limit, the probability to have a single spiral core is too low to obtain reliable data points.

\begin{figure}
  \includegraphics[width=.42\textwidth]{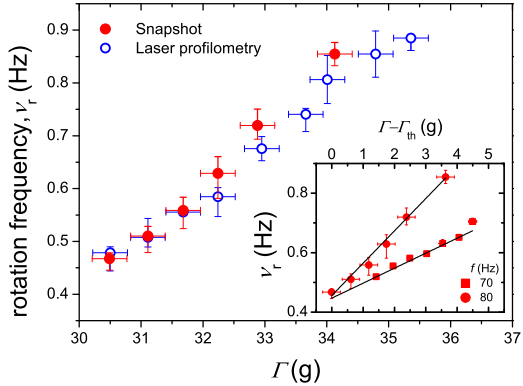}
   \caption{\label{cmp} A comparison between the rotation frequency $\nu_{\rm r}$ obtained from top view snapshots and that from the laser profilometry method for $f=80$\,Hz. Inset shows the dependence of $\nu_{\rm r}$ on the deviation of $\Gamma$ from the threshold value $\Gamma_{\rm th}$ obtained from top view snapshots for both $f=70$\,Hz and $80$\,Hz. Solid lines are linear fits to the data.}
\end{figure}

Since the rotation speed of spiral arms is stable with time, a comparison of the results obtained with both methods described above is possible. Figure~\ref{cmp} shows such a comparison for $f=80$\,Hz and various $\Gamma$, obtained by collecting both $g(t)$ and $h(t)$ information consequently for each run of experiment. The error bars correspond to the variation of $\nu_{\rm r}$ for different runs of the experiment. The good agreements between the results from both methods indicate that the dependence of $\nu_{\rm r}$ on $\Gamma$ obtained does not rely on the method chosen, thus either method can act as a counterproof of the other. Technically speaking, the analysis of the top view snapshots, although it provides more straightforward information, will cost much more computing power compared with the analysis based on laser profilometry method, because the latter contains only one dimensional information. Therefore the laser profilometry method will be the better candidate for the future analysis on a wide range of control parameters.

In the inset of Fig.~\ref{cmp}, the initial rotation frequency at the threshold acceleration $\nu_{\rm c}(\Gamma_{\rm th})$ and the slope of $\nu_{\rm r}$ with $\Gamma$ for both frequencies are compared. The dependence of $\nu_{\rm r}$ on $\Gamma$ obtained so far suggests a linear dependence. Linear fits of the dependence on $\Gamma-\Gamma_{\rm th}$ yields a common threshold frequency $\nu_{\rm c}=0.45$\,Hz, and a larger slope of $\Delta\nu_{\rm r}/\Delta\Gamma=0.11$\,Hz for $f=80$\,Hz, compared to the slope $0.05$\,Hz for $70$\,Hz.

In conclusion, the rotation dynamics of spiral pattern in vertically agitated wet granular layer are characterized with two methods: one based on the horizontal motion of the spiral arms and the other one based on the vertical height fluctuation of a single illuminated line. For the case of single spiral core, both methods reveal that the rotation frequency of spiral arms grows linearly with the peak vibration acceleration and there exists a finite initial speed at the threshold acceleration where the pattern emerges. 

Detailed analysis on a broader range of driving parameters, and for the case of multiple spiral cores will be a focus of further investigations. Since the time scale of the rotation arises presumably from the internal granular temperature gradient \cite{Huang11}, the analysis will help us to gain insight into the dynamics of fluidized wet granular matter. 

We are grateful for the support from Deutsche Forschungsgemeinschaft through HU1939/2-1.



\bibliographystyle{aipproc}   



\end{document}